\documentstyle[aaspp4,12pt]{article}
\tighten

\begin{document}

\begin{center}

\title{A Comparison of the Morphology and Stability of Relativistic
and Nonrelativistic Jets}

\author{Alexander Rosen\altaffilmark{1}, rosen@eclipse.astr.ua.edu,}
\author{Philip A. Hughes\altaffilmark{2}, hughes@astro.lsa.umich.edu,}
\author{G. Comer Duncan\altaffilmark{3}, gcd@chandra.bgsu.edu,}
\author{and Philip E. Hardee\altaffilmark{1}, hardee@venus.astr.ua.edu}

\altaffilmark{1}{Department of Physics and Astronomy, University of Alabama, Tuscaloosa, AL 35487}

\altaffilmark{2}{Astronomy Department, University of Michigan, Ann Arbor, MI 41809}

\altaffilmark{3}{Department of Physics and Astronomy, Bowling Green State University, Bowling Green, OH 43403}
\end{center}

\centerline{Accepted for publication by The Astrophysical Journal}
\centerline{To appear in the May 10, 1999 issue}

\begin{abstract}
We compare results from a relativistic and a nonrelativistic set of 2D
axisymmetric jet simulations.  For a set of five relativistic
simulations that either increase the Lorentz factor or decrease the
adiabatic index we compute nonrelativistic simulations with equal
useful power or thrust.  We examine these simulations for morphological
and dynamical differences, focusing on the velocity field, the width of
the cocoon, the age of the jets, and the internal structure of the jet
itself.

The primary result of these comparisons is that the velocity field of
nonrelativistic jet simulations can{\it not} be scaled up to give the
spatial distribution of Lorentz factors seen in relativistic
simulations.  Since the local Lorentz factor plays a major role in
determining the total intensity for parsec scale extragalactic jets,
this suggests that a nonrelativistic simulation cannot yield the proper
intensity distribution for a relativistic jet.  Another general result
is that each relativistic jet and its nonrelativistic equivalents have
similar ages (in dynamical time units, $\equiv$ R/$a_{\rm a}$, where R
is the initial radius of a cylindrical jet and $a_{\rm a}$ is the sound
speed in the ambient medium).  Also, jets with a larger Lorentz factor
have a smaller cocoon size.

In addition to these comparisons, we have completed four new
relativistic simulations to investigate the effect of varying thermal
pressure on relativistic jets.  The simulations confirm that faster
(larger Lorentz factor) and colder jets are more stable, with smaller
amplitude and longer wavelength internal variations.  However, an
exception to this occurs for the hottest jets, which appear the most
stable.  The apparent stability of these jets does not follow from
linear normal mode analysis, which suggests that there are available
growing Kelvin-Helmholtz modes.  However, these modes are not excited,
due to a lack of perturbations able to couple to them.

As an example of how these simulations can be applied to the
interpretation of observations, we use our results to estimate some
parameters of \mbox{Cygnus A}.  Although none of these estimates alone
can determine if the jets in \mbox{Cyg A} are relativistic or
nonrelativistic, estimates for the age and the jet-to-ambient density
ratio confirm values for these parameters estimated by other means.
\end{abstract}


\lefthead{Rosen, Hughes, Duncan, and Hardee}
\righthead{Comparison of Jet Simulations}

\section{Introduction}

Jet flows in extragalactic radio sources and those associated with the
Galactic superluminals could have relativistic velocities for a
significant fraction of their length  (for example, for extragalactic
sources see the arguments in Cawthorne 1991, and for Galactic
superluminals see the introduction in Hardee et al.\ 1998, hereafter
HRHD).  Recently, a number of groups have begun to simulate such
relativistic flows (\cite{dun94}; \cite{fal96};  Komissarov \& Falle
1997; \cite{mar95}; and \cite{mar97}).  There have even been
preliminary investigations of relativistic magnetohydrodynamics
(\cite{koi96}; van Putten 1996), including one attempt at combining MHD
with general relativity in 3 + 1 dimensions (Koide, Shibata \& Kudoh
1998).  While these papers and a review (\cite{nor96}) have discussed
some of the qualitative differences between relativistic and
nonrelativistic jets, a more thorough comparison is essential in order
to understand whether nonrelativistic simulations can be applied to the
interpretation of observed relativistic flows, and whether a comparison
of nonrelativistic and relativistic simulations can provide a
diagnostic of flow speed.

Many researchers have performed simulations of nonrelativistic jets
over the last 10--15 years, with reviews of the early work in
\cite{nor84} and \cite{nor85}.  Investigations frequently focused on
the question of jet disruption by fluid instabilities, and in this
light \cite{nor84} discuss the susceptibility of the jet to the
ordinary or fundamental and reflecting modes of the Kelvin-Helmholtz
(hereafter KH) pinch axisymmetric instability.  Much recent literature
uses the terms surface wave and body wave for the ordinary and
reflecting mode, respectively, and we adopt that here.  For each of the
pinch, helical, triangular, rectangular and higher order normal KH {\it
modes} there is one surface and many body {\it waves}.  Axisymmetric
simulations, like the ones discussed here, only allow waves of the
pinch mode, since higher order modes require non-axisymmetry.  Note
that the higher order modes have smaller growth lengths (or faster
growth rates) than the pinch mode, so axisymmetric jets remain
essentially stable for longer distances than in simulations where
higher order modes are allowed (e.g., a 2D slab jet simulation: Hardee
\& Norman 1988; Bodo et al.\ 1994; Bodo et al.\ 1995; or any 3D jet
simulation: e.g., Clarke 1996).  Another reason for expecting that the
flows studied here will exhibit little instability is that previous
analytic and numerical work has shown that jets with relativistic flow
speeds are more stable than jets with slower flow speeds (HRHD;
\cite{mar97}), primarily because the former jets behave as if they are
more dense.

In this paper we will compare 2D axisymmetric nonrelativistic jet
simulations performed with ZEUS-3D and simulations performed with a
relativistic Godunov (shock-capturing) scheme discussed in DH94.  For
each relativistic simulation we have computed nonrelativistic
equivalents as suggested by Komissarov (1996), based on equating the
{\it useful} power (i.e., the enthalpy flux minus the mass-energy flux)
or the thrust of the jet at the inlet.  We show that one of these sets
of nonrelativistic equivalents should have stability properties (i.e.,
spatial growth rates, etc.) that are similar to those of many of the
relativistic simulations.  These similar stability properties will lead
to similar structures within the jet: for example, small-scale features
along the jet axis that may grow to become shocks.

Also, we investigate the effect of jet temperature on the structure of
the relativistic simulations.  As opposed to the morphology of
nonrelativistic jets, which primarily depends on jet-to-ambient density
and Mach number (assuming that jet and ambient thermal pressures are
equal; Norman, Winkler, \& Smarr 1984), the morphology of relativistic
jets depends additionally on temperature as an independent parameter.
One natural consequence of a ``hot" relativistic jet, where the thermal
velocities of particles are at least mildly relativistic, is that the
classical or Newtonian Mach number ($\equiv v$/$a_{\rm j}$, where $v$
is the jet velocity relative to the laboratory frame and $a_{\rm j}$ is
the sound speed within the jet; hereafter we refer to this as the
``Mach number") is small enough that the flow is transonic (with a Mach
number between one and three).  However, linear stability analysis
suggests that the spatial growth length of the KH instability is
roughly proportional to the product of the Lorentz factor and Mach
number, and not the Mach number alone (HRHD).  Thus, a hot relativistic
flow with low Mach number should be less prone to instability, although
still not completely stable, than the Mach number alone would
indicate.  In fact, Mart\'{\i} et al.\ (1997) see this relative
stability in their hot jets, which have fewer internal features.
However, their simulations spanned a wide range of specific internal
energies with hot jets quite hot and cold jets quite cold.  We have
simulated jets with a similar range, 3--4 orders of magnitude, of
specific internal energies, but with more resolution in temperature
space and with specific internal energy density in some cases closer to
unity.  We also wish to reiterate the main point of HRHD: that
relativistic jets are not unconditionally stable -- many simulations
that have been computed contain perturbations that only weakly excite
any of the available growing modes.

In \S 2, we describe the simulations with particular emphasis on the
nonrelativistic equivalents, in \S 3, we discuss the various results
while focusing on large scale structure and the growth of the KH
instability, and in \S 4, we use some of the analysis presented here to
make estimates of some fundamental quantities (e.g., the age,
jet-to-external ratios for density and enthalpy, and the Mach number)
for the jets in \mbox{Cygnus A}, a well-studied  radio source.  In
addition, we include two Appendices; in Appendix A we derive the
relativistic form of the jet head advance speed and in Appendix B we
develop a simple model to estimate the cocoon radius in either
relativistic or nonrelativistic jets.

\section{Simulations}

For the nonrelativistic simulations, we used a cylindrical coordinate
system and the same boundary conditions as did DH94.  Specifically, the
boundary conditions allow outflow everywhere with two exceptions: at
the inlet with inflow boundary conditions and along the symmetry axis,
which has a reflecting boundary.  DH94 describe a relativistic code
that uses Adaptive Mesh Refinement (AMR), a general purpose scheme
written by Quirk (1991), to gain spatial and temporal resolution.  The
simulations reported in DH94 employed one level of refinement: the
coarse grid had 6 zones/jet radius and the refined grid had 24
zones/jet radius.  We have run the nonrelativistic simulations using
ZEUS-3D with a resolution of 12 zones/jet radius and the van Leer
linear advection scheme,  which is a combination that resolves features
of the same size as the relativistic code, as determined by examining
schlieren (density gradient) plots.  The grid extends to
41$\slantfrac{2}{3}$ jet radii along the jet axis and either 10 R or
16$\slantfrac{2}{3}$ R radially, depending on whether the relevant
relativistic simulation had a Lorentz factor below or above 5,
respectively.  The nonrelativistic simulations were run on grids of
either 500 $\times$ 120 or 500 $\times$ 200 zones.

The relativistic simulations are as described in DH94.  In order to
fill a gap in the DH94 simulations at small Lorentz factors, we include
an additional relativistic simulation that has a Lorentz factor of 2.5
and was labeled Run E in HRHD.  Run E fits between Runs A and B in the
original DH94 sequence.  We point out that HRHD adopted incorrect
values for the Lorentz factor when analyzing Runs E (2.55 instead of
2.5), B (5.5 instead of 5.0) and C (14.35 instead of 10.0) and that we
use the correct values here.  Since the correct Lorentz factors reduce
by $\lesssim$ 30\% the maximum allowable wavelengths and the
wavelengths associated with the fastest growth in the most important
modes in Runs B and C, this correction does not alter significantly
either the specific results or the overall conclusions in HRHD.

For the relativistic simulations this paper generally uses the notation
of DH94; $n$ is the proper mass density, $c$ is the speed of light,
$\beta$ is $v/c$, $\gamma$ is the Lorentz factor ($\gamma \equiv (1 -
\beta^2)^{-1/2}$), $\Gamma$ is the adiabatic index, $p$ is the internal
(or thermal) pressure, $e$ is the proper mass-energy density and is
$e=nc^2 + p/(\Gamma - 1)$, and the enthalpy is $e + p = nc^2 + (\Gamma
p/(\Gamma - 1)$).  One change from DH94 is that the mass density in the
lab frame for relativistic simulations is designated $\nu$ and not R,
which we use for the initial jet radius.  In this paper, we designate
the jet-to-ambient density ratio as $\eta$, and the jet-to-ambient
enthalpy ratio as $\eta_{\rm r}$.  We will discuss $\eta$ primarily in
the nonrelativistic simulations, but where we specifically use $\eta$
in regard to the relativistic simulations, this refers the ratio of
proper densities and is 0.1 for all of our relativistic simulations.
We use the variable $\rho$ exclusively for the mass density in the
nonrelativistic simulations.  Additionally, when comparing a similar
quantity between relativistic and nonrelativistic simulations, we
occasionally refer to each with $rel$ or $non$ subscripts.

For the nonrelativistic simulations, we need specify only two
parameters, assuming that the jet and ambient thermal pressures are
equal, as in the simulations of DH94.  The morphology of 2D
axisymmetric nonrelativistic jets depends on only $\eta$ and M, the
Mach number with respect to the sound speed within the jet (Norman,
Winkler, \& Smarr 1984).  In this paper, Mach number always refers to
this ``internal" value and we will refer to the Mach number with
respect to the ambient or external medium as either M$_{\rm a}$
($\equiv v/a_{\rm a}$) or M$_{\rm x}$ ($\equiv v/a_{\rm x}$),
respectively.  Here, the ambient medium refers to the fluid into which
the jet is propagating and the external medium refers to the fluid in
contact with the length of the jet, which can be either the shocked
ambient medium or the cocoon and is important for the development of
the KH instability.  In jets with flow speeds that are relativistic,
there is a quantity analogous to the Mach number, the relativistic or
proper Mach number, $\cal{M}$  $\equiv { {\gamma \beta}/{\gamma_{\rm s}
\beta_{\rm s}} }$, as defined in the discussion of time-independent
relativistic flows by \cite{kon80}; the subscript $s$ refers to the
sound speed (e.g., $\beta_{\rm s} \equiv a_{\rm j}/c$).  Since the
nonrelativistic simulations are scale-free and $\eta$ and M are the
main independent variables, the ambient medium density and sound speed
are set initially to unity in the ZEUS code.

We will discuss two sets of nonrelativistic equivalents, both of which
are suggested by Komissarov (1996).  One can determine $\eta$ and M for
a nonrelativistic simulation by assuming that the velocities and
pressures are equal in a relativistic simulation and its
nonrelativistic equivalent and equating {\it one} of the following
fluxes:  useful power, thrust, or mass.  We feel that the mass flux
equivalent is less interesting, because the mass flux is physically
less important than the other two fluxes and quantities that are
associated with the mass flux, such as the mass of a radio lobe, are
rarely estimated from observational data.  Thus, we have chosen not to
compute the mass flux equivalent simulations.  Equating the momentum
flux of jets in the two types of simulations leads to a more
interesting situation.  Since momentum flux and ram pressure are
equivalent within either the relativistic or nonrelativistic
formulation, there is a similar expected jet-head-to-jet velocity ratio
in both a relativistic simulation and its nonrelativistic momentum flux
equivalent.  Another interesting case involves the useful power
($\equiv \gamma^2(e+p)v - \gamma nc^2v$ in the relativistic
simulations), which is the energy flux that feeds the observed
luminosity of radio lobes.

For the useful power and thrust equating cases, we list below the
equivalent nonrelativistic mass density and Mach number in terms of the
relativistic fluid variables for each region of the flow.  Recall that
$\eta$ = $\rho_{\rm j}/\rho_{\rm a}$ and thus we need to calculate the
equivalent nonrelativistic mass density in both the jet and ambient
medium.  The relationships for $\rho$ and M in the useful power
equivalent are \cite{kom96}:

\begin{equation} 
\rho  = 2 n \gamma^2 \left( {\gamma \over \gamma + 1}
+ {\Gamma p \over (\Gamma - 1) nc^2} \right) 
\end{equation}

\begin{equation} 
{\rm and}~~~~{\rm M}^2 = 2 {{\cal{M}}^2} \left[ {
{\left( {\gamma \over \gamma + 1} + {\Gamma \over (\Gamma - 1)}{p \over
{nc^2}} \right)} \over \left( { 1 + { {\Gamma(2 - \Gamma)} \over
{\Gamma - 1} }{p \over {nc^2}} } \right) }\right].  
\end{equation}

\noindent  The relationships that follow from equating thrust are:

\begin{equation} 
\rho = \gamma^2 n \left(1 +  {\Gamma \over {\Gamma -1}} 
{p \over {n c^2}} \right) 
\end{equation}

\begin{equation} 
{\rm and}~~~~{\rm M}^2 = (\gamma^2 - 1) \left({ {n
c^2} \over {\Gamma p} } + { 1 \over {(\Gamma - 1)} } \right).
\end{equation} 

\noindent Note that equation (4) reduces to M$_{\rm non}$ =
$\gamma$M$_{\rm rel}$.  Of the relativistic simulations, the pressure
in the ambient medium is most significant in Runs C and D, increasing
the equivalent $\rho$ and decreasing $\eta$ by a factor of three (for
useful power) or two (for thrust).

We will refer to each of the nonrelativistic equivalents by the
relativistic simulation to which they are similar and two letters
describing the nature of the equivalence, e.g., B$_{\rm pw}$ and
B$_{\rm th}$ for the useful power and thrust equivalent, respectively,
for the relativistic simulation Run B.  Since we computed only one
nonrelativistic equivalent of Run A, we will refer to this as Run
A$_{\rm pw}$.

The values of $\eta$, internal Mach number (M) and ambient Mach number
(M$_{\rm a}$) are shown in Table 1 for both sets of nonrelativistic
simulations.  Note that Run D in DH94 used a softer adiabatic index,
$\Gamma = \slantfrac{4}{3}$, for the jet and ambient medium, and that
we have reproduced this in the nonrelativistic equivalents of Run D.
Since our relativistic simulations have the same $\eta$, for these
simulations we list the enthalpy ratio $\eta_{\rm r}$, which is one
measure of relativistic thermal motions, in Table 1.  The progression
of $\eta_{\rm r}$ is related to the increase in $p$ as $\gamma$ is
increased in the set of relativistic jets.  Since these simulations
model light jets pressure-matched with the ambient medium, increasing
$p$ leads to $\eta_{\rm r}$ asymptotically approaching unity from
below.

The nonrelativistic equivalent jets are frequently denser than the
external medium, which is a consequence of relativistic jets behaving
as if they have a higher density than their proper density.
Specifically, $\eta$ would have to be $\lesssim 10^{-3}$ in the most
relativistic simulations (Runs C and D) for nonrelativistic equivalents
to have $\eta <$ 1.  Examining the values of M$_{\rm a}$ reveals that
the most relativistic simulations have power- and thrust-equating
nonrelativistic equivalents that are transonic (1 $\lesssim$ M$_{\rm a}
\lesssim$ 3) with respect to the ambient medium.

One goal of this analysis is to compare the analytically predicted
growth of the KH instability modes with the internal structure found in
the two sets of simulations.  We can derive Mach numbers and
jet-to-external (i.e., cocoon, {\it not} ambient) density ratios for a
nonrelativistic jet with stability properties similar to those of a
relativistic jet by considering the approximation for the wave speed of
KH-induced disturbances at resonance (or maximum growth), $v_{\rm
w}^{*}$, for a relativistic cylindrical jet (from HRHD, equation 6c):
\begin{equation} v_{\rm w}^{*} \approx \frac{ \gamma [{\rm
M}^2-\beta^2]^{1/2} } { [{\rm M}_{\rm x}^2-\beta^2]^{1/2}+\gamma [{\rm
M}^2-\beta^2]^{1/2} }v .  \end{equation} The product $\gamma$M in
relativistic jets appears approximately in the same role as the Mach
number in nonrelativistic jets, and is the same form used for M in the
thrust-equating case.  In order to estimate the jet-to-external density
ratio, $\eta_{\rm x}$, we take the nonrelativistic limit for $v_{\rm
w}^*$/$v$: M/(M$_{\rm x}$ + M).  Assuming that the jet and external
pressures are equal, this term reduces to $\sqrt{\eta_{\rm
x}}/(\sqrt{\eta_{\rm x}} + 1)$.  Using values of $\beta$, M, and
M$_{\rm x}$ from the relativistic simulations, we solve for desired
values of $\eta_{\rm x}$ in the nonrelativistic simulations by equating
$v_{\rm w}^*/v$ in eq.\ (5) to its nonrelativistic limit.  Note that
typically M$_{\rm x} \ne$ M$_{\rm a}$, but that we do list M and
M$_{\rm x}$ for Runs E, B, and C below.  For the nonrelativistic
equivalents of all but relativistic Run A, $\eta_{\rm x}$ corresponds
to an overdense jet (see Table 1), dense enough that there is no
significant difference between $\eta_{\rm x}$ and $\eta$.
Additionally, these values are roughly similar to the set of density
ratios used in the thrust-equating simulations, which thus have {\it
both} independent parameters appropriate for matching KH stability
properties.  Even though Runs E$_{\rm th}$ and B$_{\rm th}$ have
density ratios quite different from the desired $\eta_{\rm x}$, the
density ratios in Runs C$_{\rm th}$ and D$_{\rm th}$ are a closer
match.  Also, note that the $\eta$ in Run B$_{\rm th}$ is similar to
the desired $\eta_{\rm x}$ for Run E, with only a small difference in
Mach number between E$_{\rm th}$ and B$_{\rm th}$.  This suggests that
the jet in Run B$_{\rm th}$ should have similar stability properties as
those of the jet in relativistic Run E.

\section{Results}

We show schlieren plots, in which darker regions indicate a larger
density gradient, for all of the simulations as the bow shock nears the
right edge of the grid ($z \sim$ 40 R) in Figure 1.  Typical jet
features include a bow shock, a terminal shock or Mach disk of the jet
flow, a cocoon of material lighter than the jet that has passed through
this shock, and internal structure in the form of biconical shocks
along the jet.  In the smaller Lorentz factor relativistic simulations,
there is some curvature of the terminal shock, which is the very thin
line close to the leading edge of the jet and almost perpendicular to
the jet axis in the schlieren images.  There is some difference in this
curvature between the simulations, particularly between Runs E and B,
and in general the shape and curvature of the terminal shock varies
with time.  In some of the simulations and in particular Run C, there
is a secondary shock roughly 10--20R behind and parallel to the bow
shock, which is caused by shocked ambient material expanding
supersonically backwards.  This secondary shock is weaker, or
nonexistent in Run D, a consequence of slower expansion accompanying
the softer adiabatic index.  Some of the schlieren plots show a
nonphysical reflection of the bow shock at the outer radial boundary,
generated by the edge of the numerical grid.  The strength of this
feature is somewhat exaggerated in a schlieren plot and the reflected
bow shock does not interact significantly with any of the more
important jet structures.  Despite the appearance that the reflection
coincides with an enhancement in cocoon width in Run B, we have
determined that these coincident structures are unrelated.  In this
case, the expansion of the cocoon occurs before the reflection
arrives.

From the schlieren images in Figure 1, we see that the morphology of
the jets changes with Lorentz factor in a similar manner for all three
sets of simulations.  However, the two codes do not give identical
results for the relativistic simulation with a nonrelativistic jet flow
speed and for its nonrelativistic equivalent, i.e.\ Runs A and A$_{\rm
pw}$.  While the cocoon width is similar, the detailed structure of the
cocoon with the relativistic code is more regular and appears somewhat
underesolved.  This is caused by the relatively poor capturing of
contact discontinuities by the Relativistic Harten-Lax-van
Leer-Einfeldt (RHLLE, see Einfeldt 1988 and DH94) solver used in the
relativistic simulations.  These differences within the cocoon lead to
the less regularly spaced biconical shocks within the jet in the
nonrelativistic equivalent (Run A$_{\rm pw}$).  Given these differences
between the numerical schemes for the same simulation, we will focus
primarily on large-scale differences such as in the cocoon, jet head,
and bow shock between a relativistic simulation and its nonrelativistic
equivalents.

\subsection{Cocoon}

\subsubsection{Morphological Differences}

With the aim of examining any differences in the appearance between
relativistic simulations and their nonrelativistic equivalents and
since in the nonrelativistic simulations we can rescale the velocity
field, we do so by equating inlet velocities for a relativistic
simulation and its nonrelativistic equivalent.  From the rescaled
$\beta$ in each zone in the nonrelativistic simulations, we compute a
scaled Lorentz factor.  Doppler boosting is to first order the primary
contributor to the observed intensity from the flow.  Because Doppler
boosting dominates variations in the inferred intrinsic emissivity, for
a given angle of view the Lorentz factor should provide a useful (if
coarse) estimator of the appearance of the jets at radio wavelengths.

We display contours of two values of Lorentz factor for the
relativistic simulations and for the same two values of scaled Lorentz
factor for the useful power set of nonrelativistic simulations in
Figure 2.  The adopted contours in these plots are at a low Lorentz
factor, which is near unity but different for each simulation (see
Fig.\ 2 for the values), and at the inflow Lorentz factor in the
appropriate relativistic simulation.  Assuming that the fluid flow is
along the jet and that an observer is viewing the source close to the
critical cone (with a viewing angle $\approx$ 1/$\gamma$), the radial
width of these contours suggests an effective radius of observed
intensity for the flow.  Since we will not compare simulated intensity
{\it within} any set of simulations, assuming different viewing angles
for each relativistic simulation does not invalidate the following
analysis.  We show only the useful power nonrelativistic equivalents in
Figure 2 because the thrust-equated set has a similar appearance.

Since nonrelativistic hydrodynamics imposes no upper bound on velocity,
rescaled velocities above $c$ are unavoidable.  Rescaled $\beta$s above
unity occur in most of the nonrelativistic equivalents of Runs C and D
(the only exception of the four cases is Run D$_{\rm pw}$), although
the maximum rescaled $\beta$ in any simulation is 1.02 in Run C$_{\rm
th}$.  Therefore, these equivalents are able to accommodate velocity
rescaling reasonably well.  The maxima of the rescaled $\beta$ is
smaller for equivalents of Run D than for those of Run C.
Correspondingly, the maximum speed in relativistic Run D is smaller
than the maximum speed in relativistic Run C; one effect of a softer
adiabatic index is to reduce the acceleration of the jet.

In general, the relativistic simulations have a much larger width as
measured by these Lorentz factor contours than do the nonrelativistic
simulations.  In Runs A and E, there are only small regions of high
$\gamma$ along the jet axis.  This is also true for Run A$_{\rm pw}$,
but not for Run E$_{\rm pw}$.  This similarity between Run A and
A$_{\rm pw}$ suggests that the different numerical methods admit
similar accelerations in the simulations that are most similar (note
that a $\gamma$ of 1.05 represents an acceleration of less than 2\%
above the initial jet velocity in Run A), even though the appearance
differs considerably in detail.  While some aspects of the contours in
Runs A and A$_{\rm pw}$ are similar, the most similar contours of any
relativistic-nonrelativistic pair shown in Figure 2 are the contours in
Runs D and D$_{\rm pw}$, which are narrow over most of the jet length
and only display a small cocoon-like bulge near the terminal shock.
Also, noticeable in the high-$\gamma$ contour is the pressure wave
close to the inlet, which was discussed in HRHD (the contour starts at
$r \sim$ R near the inlet and moves inward farther along the jet
axis).  This pressure wave is easiest to see in Runs C, D, and D$_{\rm
pw}$, and is perhaps seen in Runs B and E$_{\rm pw}$ as well.

Given our assumption that there is a correspondence between the Lorentz
factor and the perceived radio intensity from simulated jets, we see
from Figure 2 that the shape of emission regions are quite different
between the pairs of simulations.  Specifically, high Lorentz factor or
``bright" regions are smaller near the jet head and the internal
structure is more varied in the nonrelativistic equivalents than in
their relativistic counterparts.  These differences between
relativistic and nonrelativistic equivalents may also be seen in
lighter jets (Komissarov \& Falle 1996), but there the cocoon in the
relativistic simulation contains significant backflow, suggesting that
the cocoon could be Doppler enfeebled and much dimmer than the jet flow
if the jet is directed toward the observer.  The Doppler enfeeblement
of radiation from a cocoon with a significant backflow could dim the
cocoon by a large enough amount that light relativistic jets appear
very narrow when the jet is directed nearly toward the observer.  This
narrow appearance of the cocoon and jet from a relativistic flow could
reduce the difference between the apparent lobe widths of relativistic
jets and their nonrelativistic equivalents.

In both relativistic and nonrelativistic sets shown in Figure 2, the
simulations with the most contours crossing the jet axis are for jets
that have an underlying flow with a moderate $\gamma$, between roughly
2 and 5.  Although this comparison only shows contours of $\gamma$ and
is not a true representation of the total intensity from a radio
source, this is highly suggestive that jet flows of moderate Lorentz
factor may lead to the most ``knotty" appearance, which confirms a
result of \cite{mio97}.

We conclude that, despite the modest success of predicting the maximum
flow velocity when rescaling a nonrelativistic simulation to the
relativistic inflow speed, the rescaled velocities underscore the
necessity of performing fully relativistic hydrodynamic simulations:
the entire velocity field computed by a nonrelativistic simulation
cannot be simply rescaled to yield the velocity field of a relativistic
jet.

\subsubsection{Comparison with Simple Models}

By comparing the width of the cocoon seen in simulations with analytic
estimates based on simple assumptions, we determine the viability of
such an analytic model for the interpretation of data, and gain insight
into the origin of different flow speed-dependent morphologies.  We
derive equations for the ratio of cocoon to jet radius (R$_{\rm c}$/R)
for each of the relativistic and nonrelativistic cases in Appendix B.
From these equations, we estimate a cocoon radius based on the
relativistic form (equation [6]) for each of the relativistic
simulations.   While this relation is most applicable to Run D, because
the derivation assumes $\Gamma$ = 4/3, we expect that the different
adiabatic index in Runs A, E, B, and C will have little effect on the
cocoon radius and that it is valid to generalize this simple model to
include other adiabatic indices.  This follows from the results of
Duncan, Hughes, \& Opperman (1996), who recomputed Run C using a
modified version of the relativistic hydrodynamic code that allowed for
a variable adiabatic index.  The recomputed simulation demonstrated
that even though the simulation displayed significant spatial and
temporal variations in $\Gamma$, there were minimal differences in the
flow morphology.

We display contours of R$_{\rm c}$/R in Figure 3 from the analytic
model in $\eta_{\rm r}$---$\gamma$ space for the relativistic case from
Eq.\ 6 (derived in Appendix B.1)   and in $\eta$---M space for the
nonrelativistic case from Eq.\ 7 (derived in Appendix B.3):

\begin{equation}
\left(\frac{R_{\rm c}}{R_{\rm j}}\right)^2=1+\left(\frac{p_{\rm h}}{p_{\rm j}}\right)^{1/\Gamma}
\frac{n_{\rm j}}{n_{\rm h}}\left(\frac{1+\eta_{\rm r}+2\sqrt{\eta_{\rm r}}\gamma_{\rm j}}{\eta_{\rm r}}\right)
^{1/2},
\end{equation}

\begin{equation}
\left(\frac{R_{\rm c}}{R_{\rm j}}\right)^2=1+\left(1+\eta^{-1/2}\right)\frac{\rho_{\rm j}}
{\rho_{\rm h}}\left(\frac{p_{\rm h}}{p_{\rm j}}\right)^{1/\Gamma}.
\end{equation}

\noindent For both cases, we show the positions of the appropriate
simulations in parameter space.  We have not placed the nonrelativistic
equivalents for Run D in Fig.\ 3b, since the contours are shown for
$\Gamma$ = 5/3.  However, there is only a small difference between the
expected cocoon radius between the two adiabatic indices, with the
difference being less than 20\% for any of our simulations.

We display the theoretical values of R$_{\rm c}$/R for each simulation
as vertical lines in plots of radial profiles of density and pressure
(Figure 4), which have been measured at the position of local minimum
density that is indicated by an arrow in each of the schlieren images
in Figure 1.  In the panels that show data from a relativistic
simulation, both the lab frame density and proper density radial
profiles are shown.  In all three simulation sets, the cocoon radius
expected from the simple model typically decreases as the Lorentz
factor is increased, with the exception of both nonrelativistic
equivalents of Run C, which have an expected R$_{\rm c}$ slightly
larger than that of each Run B equivalent.  Frequently the theoretical
values in Figure 4 are coincident with a local minimum in density or
pressure, usually well within the cocoon, which we interpret as the
cocoon center.  The pressure usually increases outward from this radial
position, consistent with a nearly fully expanded cocoon.   While all
three data profiles typically are similar within each relativistic
simulation, there is one exception.  The expected R$_{\rm c}$ in Run C
occurs at a smaller radius than the position of the minimum proper
density and a larger radius than the position of the minimum lab frame
density, but it is well-centered within a wide minimum in pressure.
Thus, the analytic estimate of R$_{\rm c}$ is a better indicator of the
radial position of minimum pressure.  Also, this estimate is closer to
local extrema in the relativistic cases than in the nonrelativistic
simulations.  Specifically, in the nonrelativistic equivalents of Runs
C and D, the expected R$_{\rm c}$ is within the shocked ambient medium
or farther from the jet axis than any cocoon-like feature in the radial
profile.

Some of the underlying assumptions for the analytic expressions in
Appendix B are demonstrably poor.  Specifically, two of these
assumptions are unlikely to be satisfied:  1) pressure balance between
the jet, cocoon, and the ambient medium and 2) a slow advance speed of
the terminal shock in the lab frame.   The radial profiles in Figure 4
show that the variation of pressure is typically an order of magnitude
between the jet and the cocoon, demonstrating the weakness of the first
assumption.  The larger effective density of relativistic jets suggests
a jet head advance speed nearly that of the jet flow speed (based on
the relation for $\beta^*_{\rm h}$ in eq.\ [A10]), which would violate
the assumption of a slowly moving jet head.  This is confirmed by
comparing the average velocities of the terminal shock,
$\bar{\beta_{\rm h}}$ in Table 2, with $\beta \gtrsim$ 0.9$c$ for Runs
E, B, C, and D ($\beta$ = 0.3 in Run A).  The approximation of a slowly
moving head is poorly satisfied for all simulations but Runs A or
A$_{\rm pw}$.  Nevertheless, we see that the predicted cocoon radii are
close to those simulated, which suggests that the overall simple model
is a valid framework in which to interpret the results.

The analytic expressions for R$_{\rm c}$/R successfully mimic the
typical reduction in cocoon size that accompanies the increase in
Lorentz factor in the relativistic runs and for the most part their
nonrelativistic equivalents.  Primarily because our sets of simulations
do not have a large range in estimated R$_{\rm c}$/R, this estimate is
not a good discriminate between a relativistic and nonrelativistic flow
in our simulations.  Nevertheless, we see that either a small $\eta$
with moderate Mach number or a small $\eta_{\rm r}$ with moderate
Lorentz factor is required to generate the typically large cocoon of a
FR II source.

\subsection{Jet Head}

Balancing the sum of ram plus thermal pressure at the jet head between
the jet and the ambient medium leads to a predicted jet head advance
speed, which we will designate $\beta_{\rm h}^*$ and is equal to
${\sqrt{\eta_{\rm r}}}\beta/(\sqrt{\eta_{\rm r}} + 1/\gamma)$.  In
Appendix A, we show a slightly different method for deriving this than
given by Mart\'{\i} et al.\ (1997), clarify the approximations used,
and introduce notation used in Appendix B.  We define the efficiency as
the mean velocity divided by this predicted jet head velocity,
i.e.\ $\bar{\beta_{\rm h}}$/$\beta_{\rm h}^*$.  We list the advance
speeds and the efficiency for each simulation in Table 2, where we have
again used inlet velocities to rescale all velocities in the
nonrelativistic simulations. From the data in Table 2, it can be seen
that the efficiencies in the nonrelativistic equivalents are within
25\% of the efficiencies in the relativistic simulations.  Since the
only inaccuracy in the position of the terminal shock is the numerical
and artificial viscosities that smear the shock over a few zones, the
differences in efficiencies between a relativistic and nonrelativistic
pair is larger than the error in the measured position.

There is a smaller difference between the efficiencies of these
relativistic simulations and those of each run's nonrelativistic
equivalent than between the efficiencies of relativistic simulations in
Mart\'{\i} et al.\ (1997) and those of nonrelativistic simulations of
cylindrical jets in Norman, Winkler, \& Smarr (1983).  The main reason
for this smaller difference is that these two sets of simulations have
dissimilar initial conditions.  Specifically, efficiencies as large as
1.25 in Mart\'{\i} et al.\ (1997) were associated with highly
supersonic, cold jets with an adiabatic index of 4/3, which is unlikely
to be equivalent in any sense (e.g., equal in thrust, power, or
adiabatic index) to the nonrelativistic simulations in Norman, Winkler,
\& Smarr (1983) or those presented here.  The hot relativistic jets in
Mart\'{\i} et al.\ (1997) have efficiencies near unity, similar to
those of our hot relativistic Runs C and D.

Our simulations generally have efficiencies larger than the smallest of
those in Norman, Winkler, \& Smarr (1983), who point out that for
nonrelativistic jets of equal $\eta$ the efficiency is at a minimum for
Mach numbers of roughly 3.  The nonrelativistic equivalent simulations
have Mach numbers between 6.0 and 25.4, so the efficiencies in our set
are understandably larger than the smallest in Norman, Winkler, \&
Smarr (1983), which had an efficiency of 0.49.  Of our nonrelativistic
simulations, Run A$_{\rm pw}$ has the smallest Mach number and
efficiency of our nonrelativistic simulations, and it is also one of
the Norman, Winkler, \& Smarr (1983) set; for the efficiency that we
list as 0.61, they find 0.59.  Thus, we see that our relativistic
simulations have efficiencies smaller than those in Mart\'{\i} et
al.\ (1997), and our nonrelativistic simulations have efficiencies
larger than in Norman, Winkler, \& Smarr (1983).  Despite the
implication of Mart\'{\i} et al.\ (1997) that only hypersonic (M
$\gtrsim$ 30) nonrelativistic jets have efficiencies similar to those
of relativistic jets, we are able to match the efficiencies of
relativistic jets with our set of power- or thrust-equivalent
supersonic, dense jets.

As with the simulations of Mart\'{\i} et al.\ (1997), we find that jets
with higher Lorentz factors and their nonrelativistic equivalents have
greater efficiencies, which is related to the ability of faster,
effectively denser jets to have a narrow jet head (see Figure 1).
Since the derivation for $\beta_{\rm h}^*$ assumes that the areas on
both sides of the terminal shock are equal, the degree of collimation
of the jet flow is a primary cause for differences between estimated
and computed jet head speeds in any of the simulations.  In fact, an
explanation for efficiencies greater than unity in the simulations of
Mart\'{\i} et al.\ (1997) is that the softer adiabatic index in a {\it
cold} cylindrical relativistic jet can focus the jet flow and
accelerate the terminal shock.  In comparing the efficiencies of Run C
and D, we see that the smaller adiabatic index in a hot relativistic
jet does not significantly affect the efficiency.

\subsection{Bow shock}

The useful power and thrust equating cases for Runs C and D display bow
shocks that are weaker in the schlieren images than for the other
simulations because the jet is transonic with respect to the ambient
medium (see M$_{\rm x}$ in Table 1).  Although the strength of the bow
shock is different between the relativistic and nonrelativistic
simulations, bow shocks are not usually identifiable in extragalactic
radio sources. Indeed, the faintness of bow shocks has been suggested
by radiative transfer calculations using these relativistic simulations
(Mioduszewski, Hughes, \& Duncan 1997).  We conclude that this
difference cannot be used to distinguish relativistic jet flows in
actual sources.  However, the bow shock can affect the evolution of the
cocoon via the shocked ambient medium.  For example, a larger $\cal{M}$
reduces the Mach angle between the bow shock and the jet axis and leads
to greater confinement of the cocoon via an enhanced thermal pressure
in the shocked ambient medium.

We list the ages of the jets (in units of a dynamical time, t$_{\rm
dyn} \equiv$ R/$a_{\rm a}$) in Table 2.  In general, the ages for
nonrelativistic equivalents are similar to those for the corresponding
relativistic simulation.  Of particular note is the closeness in ages
of the thrust equating cases with the relativistic simulations, which
is related to the equivalence of momentum flux and ram pressure, as
discussed in \S 2.  Note that a comparison of ages in dynamical times
is more relevant between a relativistic simulation and its
nonrelativistic equivalents than within any set of simulations, since
the concept of a fiducial ``dynamical time" is most useful for cases
where R and $a_{\rm a}$ can be scaled to the same value in each
compared simulation.  This cannot be true for this relativistic set of
simulations because the relativistic equations require that velocities
are not scale-free (and the temperature in the ambient medium is
different for each of these relativistic simulations), which is not a
requirement of the nonrelativistic simulations.  Larger temperatures
and sound speeds in the relativistic simulations lead to smaller
dynamical times, and explain the larger ages in the hotter and faster
relativistic simulations.

\subsection{Internal Structure}

\subsubsection{Comparison of Instability in Relativistic and Nonrelativistic Simulations}

Motivated by the result of HRHD that a larger Lorentz factor is
associated with a smaller variation of pressure along the jet axis, we
analyze these pressure variations in the nonrelativistic simulations.
HRHD pointed out that the variations in Runs A, E, and B are typically
larger than those of Runs C and D.  Since the thrust-equating
nonrelativistic simulations and the relativistic simulations should
have similar stability properties (e.g., similar wave speeds and growth
lengths, see discussion in \S 2), we display the pressure variations
along an axial slice in both sets of simulations in Figure 5.  These
slices are taken from a row of zones near r = R/8 to avoid
uncharacteristically large deviations that occur along the symmetry
axis in a cylindrical jet.

A comparison of the axial pressure slices reveals that there are fewer
and smaller deviations along the sequence for the nonrelativistic
simulations as well.  In addition, in the thrust-equating cases we
reproduce the fairly abrupt break between many large-amplitude
variations in Runs A, E, and B and the few small-amplitude variations
in Runs C and D.   Recall that the analysis in \S 2 suggests that the
stability properties of Run B$_{\rm th}$ is most similar to that of Run
E of the relativistic set.  Since we have grouped the simulations into
two groups between Runs B and C, this distinction is not very useful.
The trend of fewer and smaller variations in pressure for faster jets
is also evident in axial slices of the rescaled axial velocity and to
some extent in the rescaled radial velocity, neither of which is
shown.

Although we do not display the useful power-equating set of
simulations, the set has similar behavior and demonstrates the general
trend toward smaller pressure variations as the Lorentz factor is
increased in the relativistic set.  Thus, the amplitudes of pressure
variations within the length of the jet vary similarly in both
relativistic simulations and their nonrelativistic equivalents, and the
behavior of both sets may readily be understood through linear
stability analysis.  However, the match between the stability
properties of relativistic and nonrelativistic equivalent simulations
is less than perfect.  Perhaps most importantly, for the
nonrelativistic equivalents of the largest Lorentz factor simulations
(Runs C and D), variations of the rescaled velocity are larger than the
velocity variations in the relativistic simulations because the
nonrelativistic simulations have no upper limit on their speed.  This
difference re-emphasizes the need for fully relativistic simulations.

\subsubsection{Comparison of Instability in Hot and Cold Relativistic Jets}

A detailed stability analysis of Runs A, E, B, C, and D has appeared
elsewhere (HRHD) with the conclusion that relativistic jets are {\it
not} ``unconditionally stable" (Mart\'{\i} et al.\ 1997) but that the
simulations which show jets that appear stable have not excited any
instabilities.  For example, HRHD interpret the apparent relative
stability of Run C as due to only a weak coupling of the first body
wave of the pinch mode with a pressure wave that originates at the
inlet, while the lack of cocoon vorticities prevents any perturbations
to the jet surface from jet material that has passed through the
terminal shock.  Note that this is a common mechanism for exciting the
KH instability in supersonic jets.

Ferrari, Trussoni, \& Zaninetti (1978) state that small wavelength
perturbations are ineffective for $\gamma >$ (D/$\alpha$R)$^{1/2}$
while large wavelength perturbations are stable, where D is the length
of the jet (e.g., the distance between the parent galaxy and a hot
spot) and $\alpha$ is a scaling parameter between 0.01 and 0.1.  This
confirms a general result of the stability analysis in HRHD, who found
a maximum wavelength for body waves of the pinch mode in a relativistic
jet.  The condition on $\gamma$ suggests that instabilities can grow
for ``slow" relativistic jets or within faster jets of sufficient
length.  For example, this criterion suggests that in our simulations,
which terminate when the bow shock reaches $\sim$ 40R, the jets can
show evidence of instability if $\gamma \lesssim$ 20 (if $\alpha$ is at
its maximum value), which is a condition that all of our relativistic
jets satisfy.  However, as described above, little instability was seen
due to the absence of a strong coupling of a KH mode with available
perturbations.  In comparison, the fastest simulations in Mart\'{\i} et
al.\ (1997) have $\gamma$ = 22.4 and they were computed out to only
50R, so the jet speed is sufficiently large that very little internal
structure formed. Evidently, the condition on $\gamma$ alone is
insufficient to assess the stability properties of flows, and careful
consideration must also be given to how available perturbations couple
with the available modes.

In the set of relativistic simulations, both the bulk flow (the Lorentz
factor) and the random motions (the internal pressure) were increased
in tandem:  the jets in Run A and Run E were cold, the jet in Run B was
warmer, and the jets in Runs C and D were hot.  One indicator of this
is the Mach number ({\it not} the relativistic Mach number) in each of
these simulations.  Since the sound speed within the jets varied from
0.05$c$ to nearly 0.8$c$, the Mach numbers varied between 6 in Run A
and 1.28 in Run C.  Alternatively, we can calculate the ratio of
mass-energy density to the pressure-dependent portion of enthalpy,
$(\Gamma -1)nc^2/(\Gamma p)$, which is a ratio Bicknell (1994) has
designated $\cal{R}$.  For the $\Gamma$ = 5/3 cases, $\cal{R}$ =
1/2.5$p$ (with the jet proper density and the speed of light set to
unity), which varies by an order of magnitude between each of Runs E,
B, and C, with values of 15.6, 1.4, and 0.098, respectively.  For Run D
(with $\Gamma=4/3$), $\cal{R}$ is the more familiar 1/4$p$ and equal to
0.091.  This shows that Run B is the closest to equipartition between
mass and thermal energies.  Run A is a very cold simulation with
$\cal{R} \sim$ 250.

Our simulations represent an evenly distributed sample over two orders
of magnitude in $\cal{R}$, and include a simulation near equipartition
between mass and thermal energy.  By comparison, the simulations in
Mart\'{\i} et al.\ (1997) have $\cal{R}$ of either $\cal{O}$(10) or
$\cal{O}$(0.01), but nothing in between.  Thus, our simulations
investigate a new region of parameter space for differences between hot
and cold jets.

In order to disentangle the effects of increasing Lorentz factor from
those of increasing pressure, we have completed another set of
relativistic simulations using the same code and numerical techniques
as for the first set.  These new simulations fill in places in Lorentz
factor-internal pressure ($\gamma$---$p$) space beyond Runs E, B, and C
(see Table 3).  All of the new runs have $\Gamma = 5/3$.  We name the
new simulations with the designation of an original simulation with the
same Lorentz factor and a subscript indicating a different temperature
(e.g., B$_{\rm cool}$ has $\gamma$ = 5.0 as did Run B, but a pressure
and $\cal{R}$ similar to those of the cooler Run E).  We also list the
relativistic Mach number, $\cal{M}$, and the Mach number, M, for each
case in Table 3, which shows the decrease in M that accompanies an
increase in thermal pressure.

Schlieren plots of the new set of simulations and Runs E, B, and C are
displayed in Figure 6, which has columns of equal $p$ increasing to the
right and rows of equal $\gamma$ increasing from top to bottom.  As
with the schlieren plots in Figure 1, all plots in Figure 6 are at the
time when the bow shock nears the edge of the grid.  We list the age of
each simulation in units of a dynamical time in Table 3.

These new simulations have been analyzed for their stability
properties, as was done in HRHD.  This analysis can explain much of the
internal structure in the schlieren images of Figure 6.  The normal
mode analysis suggests that faster, cooler relativistic jets are more
stable, so Run C$_{\rm cool}$ should have the least internal structure
of the simulations in Figure 6.  Any simulation in a panel to the right
of (hotter than) or above (slower than) Run C$_{\rm cool}$ should be
more unstable and have more structure.  Hotter simulations should be
more unstable because the increased sound speed shortens a jet crossing
time for a perturbation.  Faster flows are more stable because they
naturally increase the spatial growth length.  With the exception of
the hottest simulations, we do see the expected instability trends,
with the slow, hot Run E$_{\rm warm}$ appearing the most internally
unstable.  Therefore, we confirm the most basic results of the
stability analysis.

The stability analysis of HRHD also suggests an explanation for the
relative lack of structure in the ``hot" simulations.  Run B$_{\rm
hot}$ appears to have the least structure and as we have already
discussed Run C has only a few small internal variations in pressure.
Not only do the conditions in the two hot simulations admit growing
modes, as shown by the normal mode analysis, but both have external
Mach numbers, M$_{\rm x}$, near 1.3 (see Table 3).  The stability
analysis suggests that this is near a critical value, where the
solution to the dispersion relation changes dramatically and quickly.
We display pinch mode solutions from the linear analysis for the three
simulations with $\gamma$ = 5, i.e.\ Runs B$_{\rm cool}$, B, and
B$_{\rm hot}$, in Figure 7.  There is a noticeable difference between
the shape of the growth rate of the surface wave of the pinch mode for
Run B with M$_{\rm x}$ =  1.56 and for Run B$_{\rm hot}$ with M$_{\rm
x}$ = 1.31, and this difference exists even between Run C with M$_{\rm
x}$ = 1.33 and Run B$_{\rm hot}$.  As the external Mach number
decreases toward the critical value of 1.3, the growth rate of the
surface wave increases at all frequencies so that it dominates the
growth rates of all the body waves, whose solutions remain similar.  In
addition, the dominant surface wave for M$_{\rm x} \lesssim$ 1.3 does
not pass through a resonance, so there is no frequency associated with
a peak growth rate (see Figure 7).  This contrasts with the body waves,
which still have a peak growth rate while the jet is supersonic; body
waves are suppressed in a subsonic flow.  As with the hot jets in HRHD,
the stable appearance of jets in these simulations is most likely due
to a weak coupling between the instability at all frequencies and the
jet.  Additionally, some simple physical arguments for the amplitude of
displacement surfaces (e.g., as in HRHD) indicate that short wavelength
perturbations will saturate at small amplitudes.   These small
wavelengths would need adequate grid resolution to grow within a
numerical simulation.

\section{Comparison with Observations --- Cyg A}

We will now apply our results to available observations in the manner
of Williams (1991).  For the following comparison with observations, we
will concern ourselves with data for a single source --- \mbox{Cygnus
A}.  A large fraction of the observations of this source have recently
been compiled in a review article by Carilli \& Barthel (1996).  At two
orders of magnitude above the FR I--FR II break and a redshift of
0.0562 (\cite{sto94}), \mbox{Cygnus A} is a very powerful and very
close double-lobed radio source.  Based on results from a 3D MHD
nonrelativistic jet simulation with ZEUS-3D, \cite{har96} suggested
that a dense jet with $\gamma$M, which can be thought of as $\cal{M}$
in a cold jet, $\approx$ 10 and a jet-to-cocoon density ratio of
$\gamma^2\eta \approx$ 6 is needed in order to reproduce the observed
helical and filamentary structure.  A dense jet is appropriate for such
a simulation since the ambient medium represents the cocoon, which is
lighter than the jet.

While Komissarov \& Falle (1998) have modeled Cygnus
A with some success by using the self-similar properties of
an expanding jet, we apply the results from our cylindrical jets
to Cyg A by first estimating the age of the source.  
The distance of the hotspot or farthest edge of the lobe from
the core, D, is 50 h$^{-1}_{100}$ kpc, where  h$_{100}$ is the Hubble
constant normalized by $100\,{\rm km}\,{\rm s}^{-1}\,{\rm Mpc}^{-1}$.
We adopt a jet radius $\sim$ 0.5 kpc, based on the results of Perley,
Dreher, \& Cowan (1984) who observed that the jet has a constant FWHM
of 0\farcs 7 for the inner \onethird D; at the redshift quoted above
this implies a radius of 575 h$^{-1}_{100}$ pc.  Therefore the jet has
propagated $\sim$ 100 R, and it would take roughly 25--60 dynamical
times to reach that distance (based on 10--25 t$_{\rm dyn}$ to reach
$\sim$ 40 R in simulations with $\cal{M} \sim$ 8, see Table 2).  High
resolution ROSAT results indicate the inner 50 kpc of cluster has a
temperature of 3 $\times$ 10$^7$ K (Reynolds \& Fabian 1996), so the
sound speed in the external medium is roughly $800\,{\rm km}\,{\rm
s}^{-1}$ and 25--60 dynamical times corresponds to roughly 15--45
Myrs.  This age range straddles the observationally determined value of
30 Myr (Carilli et al.\ 1991), which is from a self-consistent model
based on the spectral aging of the \mbox{Cyg A} lobes.  Since the age
estimates for the nonrelativistic and relativistic are close, we cannot
use this estimate to determine whether the jet in \mbox{Cyg A}
propagates at a relativistic speed.

From the analytic expressions for the radius of the cocoon (see
eqs.\ [6] and [7]), we can also estimate $\eta_{\rm r}$ in the
relativistic case or $\eta$ in the nonrelativistic case.  The typical
radius of the cocoons in \mbox{Cygnus A} is 10 kpc, so R$_{\rm c}$/R is
roughly 20.  From eq.\ (6), assuming that the Lorentz factor is in the
range of 5--10, and accounting for the fact that the analytic
expression frequently predicted the cocoon {\it center}, we estimate an
enthalpy ratio of roughly 10$^{-4}$, with smaller Lorentz factors
requiring smaller values of $\eta_{\rm r}$ (as seen in Figure 3a).
From eq.\ (7) and assuming M is roughly 5--10, $\eta$ should be 3
$\times 10^{-5} - 3 \times 10^{-4}$. This suggestion of a very light
jet is consistent with an estimate by Clarke (1996), who from 3D
nonrelativistic simulations proposed that $\eta$ is 3 $\times$
10$^{-4}$ and M is 15.  Either of the estimates for $\eta_{\rm r}$ and
$\eta$ suggest the jet in \mbox{Cyg A} is very light, and possibly
cold.  Also, it should be noted that this value of $\eta$ is smaller
than that of any jet simulation yet attempted.

\section{Conclusions}

In this paper, we have examined morphological and dynamical differences
between simulations of relativistic and nonrelativistic jets, including
the age, efficiency, velocity field, and internal structure of the jets
as well as the size of the cocoon.  An important result of these
comparisons is that the velocity field of nonrelativistic jet
simulations can{\it not} be scaled up to give the spatial distribution
of Lorentz factors seen in relativistic simulations.  Specifically,
such a scaling substantially underpredicts the size of the region of
significant Lorentz factor.  Since the local Lorentz factor is the
primary factor determining the brightness of the source, this suggests
that a nonrelativistic simulation cannot yield the proper intensity
distribution from a relativistic jet.  From the rescaled velocity
plots, we see evidence for a result more definitively seen in
\cite{mio97}: jet flows of moderate Lorentz factor ($\gamma \sim$ 2--5)
generate the most knotty appearance.

Other general results are that relativistic simulations and their
nonrelativistic equivalents have similar ages (in dynamical time units,
$\equiv$ R/$a_{\rm a}$), efficiencies, and that jets with a larger
Lorentz factor have a smaller cocoon size.   The ages and efficiencies
in the nonrelativistic simulations with equal thrust are a particularly
good match to the relativistic simulations.  Also, for both
relativistic simulations and their nonrelativistic equivalents with
$\Gamma$ = 5/3, the widths of the cocoon radii predicted from simple
models frequently match a density minimum within the cocoon, and
therefore are adequate estimators of cocoon centers.  However, due to
the limited range of estimated cocoon widths in all of our relativistic
and nonrelativistic sets of simulations, these simple models do not
discriminate well between relativistic and nonrelativistic flow
speeds.  Slices of pressure near the symmetry axis vary similarly in
both relativistic and nonrelativistic simulation sets.  Even in the
nonrelativistic simulations, these data slices show fewer and smaller
deviations as the Lorentz factor in the equivalent relativistic
simulation is increased.

In addition to this comparison, we also have completed four new
relativistic simulations to investigate the effect of varying thermal
pressure on the structure of simulated relativistic jets.  These
simulations confirm that faster (larger Lorentz factor), colder jets
are more stable, with smaller amplitude and longer wavelength internal
variations.   However, an exception to this occurs for hot jets, where
we see evidence beyond that presented in HRHD that hot, relativistic
jets appear stable because of a weak coupling between perturbations and
available growing KH instability modes.  The stability analysis also
suggests a different behavior as the external Mach number decreases
through $\sim$ 1.3.  Jets with such a small M$_{\rm x}$ should have
large growth rates at high frequencies and small wavelengths.  These
jets appear stable also because of the weak coupling of between the
instability and the perturbations within the jet.  A higher grid
resolution may allow the smaller wavelength perturbations to grow.

As an example of how these simulations can be applied to observations,
we use our results to estimate the age, jet-to-ambient density, and
jet-to-ambient enthalpy of \mbox{Cygnus A}.  Although none of these
estimates can determine if the jet in \mbox{Cyg A} is relativistic or
nonrelativistic, they confirm independent estimates of these
parameters.  Our results suggest that the jet is light with $\eta
\approx 10^{-4}$, and possibly cold with $\eta_{\rm r} \approx 10^{-4}$
if the jet has a relativistic flow speed.

Future work will focus on extending the techniques used in the
relativistic jet simulations (i.e., AMR and RHLLE) to allow for high
resolution three dimensional relativistic simulations and the analysis
thereof.  Since the growth rates of nonaxisymmetric modes in
nonrelativistic jets are larger than the growth rates of the symmetric
modes, nonaxisymmetric relativistic jets should be more unstable than
the relatively stable simulations presented here.  Thus, investigating
the stability of 3D relativistic jets will reveal much about the nature
of astrophysical jets.

\acknowledgments{We thank the Laboratory for Computational Astrophysics
for providing the version of ZEUS-3D used for the nonrelativistic
simulations, which were run on a Sparc 10 in the Department of Physics
and Astronomy at the University of Alabama.  New simulations with the
DH94 relativistic code were performed on a Power Challenge at the Ohio
Supercomputer Center and on an Ultrasparc at the University of
Michigan.   AR and PEH wish to acknowledge support from the National
Science Foundation grants AST-9318397 and AST-9802955 to the University
of Alabama and PAH acknowledges support from the National Science
Foundation grant AST-9617032 to the University of Michigan.}

\appendix

\section{Head Advance Speed}
The momentum `discharge' of a jet is
\begin{equation}
Q=\left[\left(e+p\right)u^2+p\right]A,
\end{equation}
where $e$ is total internal energy density, $p$ is pressure, $u$ is 
the four-velocity, and $A$ is the area of the flow. An estimate of the advance
speed of the head of a jet may be made by balancing the discharges associated
with the jet and ambient media in the frame of the advancing Mach disk-contact
surface-bow shock structure. Since we assume the area is common, 
we have
\begin{equation}
\left(e_{\rm j}+p_{\rm j}\right)\gamma_{\rm c}^2\beta_{\rm c}^2+p_{\rm j}=\left(e_{\rm a}+p_{\rm a}\right)\gamma_{\rm a}^2
\beta_{\rm a}^2+p_{\rm a},
\end{equation}
where we have used $v/c=u/\gamma=\beta$, and subscript $j$ refers to the jet,
while subscript $a$ refers to the ambient medium. Subscript $c$ refers to 
the jet speed measured in the advancing head frame, so that 
\begin{equation}
\beta_{\rm c}=\frac{\beta_{\rm j}-\beta_{\rm a}}{1-\beta_{\rm j}\beta_{\rm a}},
\end{equation}
and
\begin{equation}
\gamma_{\rm c}=\left(1-\beta_{\rm c}^2\right)^{-1/2}.
\end{equation}
If the jet and ambient pressures are equal, or the jet flow is hypersonic,
the above simplifies to
\begin{equation}
\left(e_{\rm j}+p_{\rm j}\right)\gamma_{\rm c}^2\beta_{\rm c}^2=\left(e_{\rm a}+p_{\rm a}\right)\gamma_{\rm a}^2
\beta_{\rm a}^2.
\end{equation}
Making the following definitions:
\begin{equation}
e=\varepsilon \rho,
\end{equation}
\begin{equation}
\eta=\frac{\rho_{\rm j}}{\rho_{\rm a}} < 1,
\end{equation}
and
\begin{equation}
\eta_{\rm r}=\eta \frac{\varepsilon_{\rm j}+\frac{p_{\rm j}}{\rho_{\rm j}}}{\varepsilon_{\rm a}+\frac{p_{\rm a}}{\rho_{\rm a}}},
\end{equation}
we see that
\begin{equation}
\eta_{\rm r}\gamma_{\rm c}^2\beta_{\rm c}^2=\gamma_{\rm a}^2\beta_{\rm a}^2.
\end{equation}
Substituting for $\beta_{\rm c}$ and $\gamma_{\rm c}$ from equations (A3)
and (A4), after some algebra we find
\begin{equation}
\beta_{\rm a}=\frac{\sqrt{\eta_{\rm r}}\beta_{\rm j}}{\sqrt{\eta_{\rm r}}+\gamma_{\rm j}^{-1}},
\end{equation}
which is the speed of ambient flow towards the head, and thus the speed of
the head through the ambient medium.  Note that the quantity $\beta_{\rm a}$ 
is the predicted head velocity, which is designated $\beta_{\rm h}^*$ in the text.  This is the relativistic generalization
of the familiar Newtonian form
\begin{equation}
v_{\rm h}=\frac{v_{\rm j}}{1+\eta^{-1/2}}.
\end{equation}

\section{Cocoon Radius}
\subsection{Relativistic Formulation}
The particle `discharge' of a jet is
\begin{equation}
P=nv\gamma A,
\end{equation}
where $n$ is the baryon density, and other quantities are as defined above.
Jet material thermalized at a Mach disk expands laterally to from a cocoon,
but the particle discharge is conserved, so
\begin{equation}
n_{\rm j}v_{\rm j}\gamma_{\rm j}A_{\rm j}=n_{\rm c}v_{\rm c}\gamma_{\rm c}A_{\rm c},
\end{equation}
where the subscript $c$ refers to the cocoon.  As $A_{\rm j}=\pi R_{\rm j}^2$ and $A_{\rm c}=\pi R_{\rm c}^2-\pi R_{\rm j}^2$ (the jet displaces 
cocoon material), we have
\begin{equation}
\left(\frac{R_{\rm c}}{R_{\rm j}}\right)^2=1+\frac{n_{\rm j}v_{\rm j}\gamma_{\rm j}}{n_{\rm c}v_{\rm c}\gamma_{\rm c}}.
\end{equation}
If we adopt the picture that shocked jet material streams beyond the Mach
disk to the vicinity of the contact surface, and expands laterally to form
the cocoon, then, until further longitudinal expansion has given the 
material a back flow velocity in the frame of the head, this material is
moving at the speed of the head --- as characterized by $\beta_{\rm a}$ and 
$\gamma_{\rm a}$ above. Thus
\begin{equation}
\left(\frac{R_{\rm c}}{R_{\rm j}}\right)^2=1+\frac{n_{\rm j}}{n_{\rm h}}\frac{n_{\rm h}}{n_{\rm c}}
\frac{v_{\rm j}}{v_{\rm a}}\frac{\gamma_{\rm j}}{\gamma_{\rm a}}.
\end{equation}
We may write
\begin{equation}
\frac{n_{\rm h}}{n_{\rm c}}=\left(\frac{p_{\rm h}}{p_{\rm c}}\right)^{1/\Gamma}=\left(\frac{p_{\rm h}}{p_{\rm j}}
\right)^{1/\Gamma},
\end{equation}
if the jet is in pressure balance with the cocoon, so that $p_{\rm c}=p_{\rm j}$.  Using the previously-derived relation for $v_{\rm a}$, the term
\begin{equation}
\frac{v_{\rm j}}{v_{\rm a}}\frac{\gamma_{\rm j}}{\gamma_{\rm a}}
\end{equation}
may be recast, and after some algebra we get
\begin{equation}
\left(\frac{R_{\rm c}}{R_{\rm j}}\right)^2=1+\left(\frac{p_{\rm h}}{p_{\rm j}}\right)^{1/\Gamma}
\frac{n_{\rm j}}{n_{\rm h}}\left(\frac{1+\eta_{\rm r}+2\sqrt{\eta_{\rm r}}\gamma_{\rm j}}{\eta_{\rm r}}\right)
^{1/2}.
\end{equation}

\subsection{Relativistic Jump Values}
The terms $p_{\rm h}/p_{\rm j}$ and $n_{\rm j}/n_{\rm h}$ are determined by the Rankine-Hugoniot
conditions at the Mach disk. The simplest case (applicable to Run D in DH94) 
is for $\Gamma=4/3$; then these ratios are functions only of $\gamma_{\rm j}$.
Let primes denote values measured in the frame of the shock discontinuity,
and subscript $1$ signify the upstream state.  Given the extreme equation of 
state, from K\"onigl (1980) we have
\begin{equation}
\frac{p_{\rm h}}{p_{\rm j}}=\frac{2\hat\Gamma_1{\cal M}_1^2-\left(\hat\Gamma_1-1\right)}
{\hat\Gamma_1+1}=\frac{4{\cal M}_1^2-1}{3}=\frac{8\gamma'_{\rm j}{}^2\beta'_{\rm j}{}^2-1}
{3},
\end{equation}
because 
\begin{equation}
\hat\Gamma=\gamma_{\rm sound}^2\Gamma=\frac{3}{2}\times\frac{4}{3}=2,
\end{equation}
and 
\begin{equation}
{\cal M}_1=\frac{\gamma'_1\beta'_1}{\gamma'\beta'|_{\rm sound}}=
\frac{\gamma'_1\beta'_1}{1/\sqrt{2}},
\end{equation}
for the adopted adiabatic index. From mass conservation
\begin{equation}
n_1\beta'_1\gamma'_1=n_2\beta'_2\gamma'_2,
\end{equation}
so that
\begin{equation}
\frac{n_{\rm j}}{n_{\rm h}}=\frac{\beta'_{\rm h}\gamma'_{\rm h}}{\beta'_{\rm j}\gamma'_{\rm j}}.
\end{equation}
For the adopted adiabatic index, $\beta'_1\beta'_2=1/3$ (for {\it any} flow
speed), and after some algebra the above equation may be recast as
\begin{equation}
\frac{n_{\rm j}}{n_{\rm h}}=\frac{\gamma'_{\rm j}}{\left(8\gamma'_{\rm j}{}^4-17\gamma'_{\rm j}{}^2+9\right)^{1/2}}.
\end{equation}
Note that both pressure and density depend on primed quantities --- those
measured in the head frame. For situations in which the head moves
slowly compared with the jet flow speed, it will be a reasonable approximation
to use the jet parameters directly.

\subsection{Nonrelativistic Formulation}
Using arguments similar to those above we have
\begin{equation}
\rho_{\rm j}v_{\rm j}A_{\rm j}=\rho_{\rm c}v_{\rm c}A_{\rm c},
\end{equation}
so that
\begin{equation}
\left(\frac{R_{\rm c}}{R_{\rm j}}\right)^2=1+\frac{v_{\rm j}}{v_{\rm c}}\frac{\rho_{\rm j}}{\rho_{\rm c}}.
\end{equation}
It follows as before that
\begin{equation}
\left(\frac{R_{\rm c}}{R_{\rm j}}\right)^2=1+\left(1+\eta^{-1/2}\right)\frac{\rho_{\rm j}}
{\rho_{\rm h}}\left(\frac{p_{\rm h}}{p_{\rm j}}\right)^{1/\Gamma}.
\end{equation}
From the nonrelativistic Rankine-Hugoniot conditions we may write 
\begin{equation}
\frac{\rho_{\rm j}}
{\rho_{\rm h}}\left(\frac{p_{\rm h}}{p_{\rm j}}\right)^{1/\Gamma} = \left[\frac{\left(\Gamma-1\right){\rm M}^2+2}{\left(\Gamma+1\right)
{\rm M}^2}\right]\times \left[\frac{2\Gamma{\rm M}^2-\left(\Gamma-1\right)}
{\left(\Gamma+1\right)}\right]^{1/\Gamma}.
\end{equation}
Note again that Mach number is determined by the upstream flow speed in
a frame moving with the shock, but that for a fast jet and slowly moving
bow, the jet speed will yield a good approximation to this.

\clearpage

\begin{figure}
\figcaption{Schlieren plots of the relativistic and both corresponding
nonrelativistic sets of simulations.  Darker shading indicates a larger
gradient in the density in the nonrelativistic sets or lab frame
density in the relativistic set.  Each set of five vertical panels
shows Runs A, E, B, C, and D or their equivalents.  Scales shown along
each axis are in units of jet radius, R.  Each panel contains an arrow
showing the axial position of radial profiles in Figure 4.}
\end{figure}

\begin{figure}
\figcaption{Contours of low and high Lorentz factors.  We show a low
and high contour of $\gamma$ for the relativistic (Runs A, E, B, C, and
D) and of rescaled $\gamma$ in the useful power (Runs A$_{\rm pw}$,
E$_{\rm pw}$, B$_{\rm pw}$, C$_{\rm pw}$, and D$_{\rm pw}$)
simulations.  The rescaling of velocity in the nonrelativistic
simulations is described in \S 3.1.1.  Note that the plots for runs C
and D and their nonrelativistic equivalents have been cropped to show
only the inner 10R on the radial axis.} 
\end{figure}

\begin{figure}
\figcaption{Estimates of cocoon radius in $\eta_{\rm r}$---$\gamma$ or
$\eta$---M space.  The contours of equal cocoon radius, which are
labeled by R$_{\rm c}$/R, are from the equations in Appendix B.  In
panel a), the region at the upper left (above the jagged contour)
corresponds to a subsonic flow and no cocoon is allowed by the model.
The placement of each simulation in either panel is determined by the
input jet data listed in Table 1.} 
\end{figure}

\begin{figure}
\figcaption{Radial slices of density and pressure for all 14
simulations, taken at the axial position of minimum density. The solid
line shows the density, $\rho$, for the nonrelativistic simulations or
the lab frame density, $\nu$, for the relativistic simulations, the
dotted line shows the pressure, and the dashed line shows the proper
density, $n$, in the relativistic simulations.  The vertical line in
each panel is at the radius given by eq.\ (6) for relativistic Runs
B--E, or eq.\ (B7) in all other cases.  The vertical axis is the same
for each row of panels, but not each column.} 
\end{figure}

\begin{figure} 
\figcaption{Slices parallel to the axis of jet pressure relative to the
inlet jet pressure, $p_{\rm b}$.  The logarithm of pressure is shown
for a row of zones near but not on the jet axis for the relativistic
({\it solid line}) jet centered on $r$ = 5/48R, which is the middle of
the third zone, and thrust-equating nonrelativistic ({\it dashed line})
jet on $r$ = 3/24R, which is the middle of the second zone.  Since the
highest spatial resolution in the relativistic simulations is 24
zones/R and the resolution in the nonrelativistic simulations is 12
zones/R, these two rows overlap in physical space.} 
\end{figure}

\begin{figure}
\figcaption{Schlieren plots of the relativistic simulations that can be
compared to show the effect of temperature (hot vs.\ cold jets).   The
top row has $\gamma$ = 2.5, the middle one has $\gamma$ = 5, and the
bottom one has $\gamma$ = 10.  Each column differs in pressure by
roughly an order of magnitude compared to an adjacent column
(increasing left to right). } 
\end{figure}

\begin{figure}
\figcaption{Real ({\it dotted lines}) and imaginary ({\it dashed
lines}) part of the wavenumber as a function of frequency from linear
stability analysis for the $\gamma$ = 5 simulations.  For Runs B$_{\rm
cool}$, B, and B$_{\rm hot}$, we display the solutions for the surface,
labeled S, and first two body waves, labeled B$_1$ and B$_2$, of the
pinch mode.  Note the different behavior for the surface wave in Run
B$_{\rm hot}$.} 
\end{figure}

\clearpage

\begin{deluxetable}{cccccccccccc}
\tablewidth{0pt}  
\tablecaption{Simulation Parameters \label{tbl1}}
\tablehead{
&&&& & \multispan3{~Power} & \multispan3{~Thrust} & {Stability}\nl
&&&& & \multispan3{\hrulefill} & \multispan3{\hrulefill} & \hrulefill\nl
\colhead{Run\tablenotemark{a}} & $\Gamma$ & $\eta_{\rm r}$ & $\cal{M}$ & $\gamma$ & \hfil $\eta$ \hfil & M & M$_{\rm a}$  
& \hfil $\eta$ \hfil & M & M$_{\rm a}$ & $\eta_{\rm x}$\tablenotemark{c} \nl
}
\startdata
A\tablenotemark{b}    &  $\slantfrac{5}{3}$ & 0.10 &  6  &  1.05&   0.11 &  6.1 & 18.1  &   0.11 &  6.0 & 18.1 \nl
E                     &  $\slantfrac{5}{3}$ & 0.11 &  8  &  2.5 &   1.03 & 10.2 &  9.9  &   0.70 &  8.3 &  9.9 & ~8.8\nl
B                     &  $\slantfrac{5}{3}$ & 0.16 &  8  &  5.0 &   6.70 & 12.6 &  4.9  &   3.96 &  9.4 &  4.7 & 41.8\nl
C                     &  $\slantfrac{5}{3}$ & 0.56 &  8  & 10.0 &  73.1  & 18.0 &  2.1  &  55.5  & 12.8 &  1.7 & 83.0\nl
D                     &  $\slantfrac{4}{3}$ & 0.57 & 15  & 10.0 &  74.4  & 25.4 &  2.9  &  57.1  & 18.0 &  2.4 & 93.1\nl
\enddata
\tablenotetext{a}{Relativistic simulations A--D were originally discussed in DH94, Run E was added for HRHD.}
\tablenotetext{b}{A single nonrelativistic equivalent of this slow jet was run with $\eta$ = 0.1, M = 6.0 and M$_a$ = 19.0.} 
\tablenotetext{c}{Estimated from setting the wave speed of the maximum growing modes (equation [5]) to its nonrelativistic
equivalent with the appropriate values from the relativistic simulations.}
\end{deluxetable}

\begin{deluxetable}{lccccc} 
\tablewidth{4.0in}
\tablecaption{Simulation Data \label{tbl3} }
\tablehead{ 
\colhead{Run} & \colhead{$\beta_{\rm s,a}$} & \colhead{t$_{\rm dyn}$}
& \colhead{$\bar{\beta_{\rm h}}$}& \colhead{$\beta_{\rm h}^*$} 
& \colhead{Efficiency} \nl
}
\startdata
A            & 0.017 & 11.0 & 0.058 & 0.075 & 0.77 \nl
\smallskip
A$_{\rm pw}$ &       & 14.0 & 0.044 & 0.072 & 0.61 \nl
E            & 0.092 & ~8.1 & 0.314 & 0.417 & 0.75 \nl
E$_{\rm pw}$ &       & ~9.8 & 0.433 & 0.464 & 0.93 \nl
\smallskip
E$_{\rm th}$ &       & ~8.5 & 0.373 & 0.419 & 0.89 \nl
B            & 0.208 & 14.7 & 0.554 & 0.652 & 0.85 \nl
B$_{\rm pw}$ &       & 12.3 & 0.646 & 0.707 & 0.91 \nl
\smallskip
B$_{\rm th}$ &       & 14.5 & 0.575 & 0.654 & 0.88 \nl
C            & 0.581 & 26.9 & 0.828 & 0.877 & 0.94 \nl
C$_{\rm pw}$ &       & 21.5 & 0.868 & 0.891 & 0.97 \nl
\smallskip
C$_{\rm th}$ &       & 26.5 & 0.840 & 0.876 & 0.96 \nl
D            & 0.418 & 19.5 & 0.845 & 0.879 & 0.96 \nl 
D$_{\rm pw}$ &       & 15.5 & 0.876 & 0.891 & 0.98 \nl
D$_{\rm th}$ &       & 19.4 & 0.852 & 0.878 & 0.97 \nl
\enddata
\end{deluxetable}

\begin{deluxetable}{lccrrrrccc}
\tablewidth{0pt}
\tablecaption{Hot vs.\ Cold ($\Gamma$ = 5/3) Jet Simulation Data}
\tablehead{
\colhead{Run} & p & $\beta_{\rm s}$ & $\cal{R}$ & $\gamma$ & $\cal{M}$ 
& \hfil M \hfil & t$_{\rm dyn}$ (R/$a_{\rm a}$) \tablenotemark{a} & \colhead{M$_{\rm x}$}\nl
}
\startdata 
E               &  0.051 & 0.275 & 7.8 &  2.5 &  8.0 & 3.33 & 11.5     & 2.86 \nl
B$_{\rm cool}$  &  0.051 & 0.275 & 7.8 &  5.0 & 17.1 & 3.56 & \phn 5.7 & 1.96 \nl
C$_{\rm cool}$  &  0.051 & 0.275 & 7.8 & 10.0 & 34.8 & 3.62 & \phn 4.6 & 1.88 \nl
E$_{\rm warm}$  &  0.277 & 0.522 & 1.4 &  2.5 &  3.7 & 1.76 & 24.1     & 1.67 \nl 
B               &  0.277 & 0.522 & 1.4 &  5.0 &  8.0 & 1.88 & 14.7     & 1.56 \nl 
C$_{\rm warm}$  &  0.277 & 0.522 & 1.4 & 10.0 & 16.3 & 1.91 & 10.5     & 1.63 \nl
B$_{\rm hot}$   &  4.096 & 0.779 & 0.1 &  5.0 &  3.9 & 1.26 & 30.8     & 1.31 \nl
C               &  4.095 & 0.779 & 0.1 & 10.0 &  8.0 & 1.28 & 26.9     & 1.33 \nl
\enddata
\tablenotetext{a} {Dynamical times of jets simulations at the time of
the image shown in Figure 6.}
\end{deluxetable}

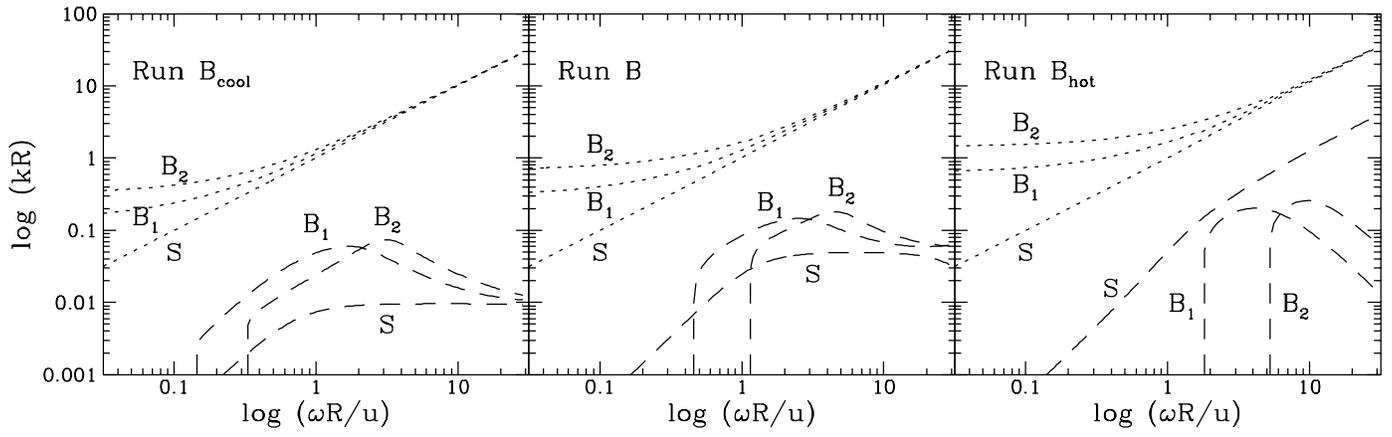
\end{document}